\documentclass[12pt,english]{article}
\usepackage{babel}
\usepackage{amsmath}
\usepackage{amssymb} 
\usepackage[small]{caption2} 
\usepackage{fleqn} 
\usepackage{graphicx} 
\usepackage[small,loose]{subfigure}  
\usepackage{cite} 
\advance \headheight by 3.0truept       
\setcaptionwidth{.85\textwidth}
\DeclareMathOperator{\re}{Re}
\DeclareMathOperator{\im}{Im}

\newcommand{\CenterObject}[1]{\ensuremath{\vcenter{\hbox{#1}}}}

\makeatother

\begin{document}
\title{
\begin{flushleft}
{\normalsize DESY 04-062\hfill April 2004}
\end{flushleft}
\vspace{2cm}
{\bf Dilaton Destabilization \\at High Temperature}\\[0.8cm]}
\author{\\
{\bf W. Buchm\"uller, K. Hamaguchi, O. Lebedev, M. Ratz} \\ \ \\ 
{\it Deutsches Elektronen-Synchrotron DESY, 22603 Hamburg, Germany}
 }
\date{}
\maketitle \thispagestyle{empty} 

\abstract{Many compactifications of higher--dimensional supersymmetric
theories have approximate vacuum degeneracy. The associated moduli
fields are stabilized by non--perturbative effects which break
supersymmetry. We show that at finite temperature the effective
potential of the dilaton acquires a negative linear term. This
destabilizes all moduli fields at sufficiently high temperature. We
compute the corresponding critical temperature which is determined by
the scale of supersymmetry breaking, the $\beta$--function associated
with gaugino condensation and the curvature of the K\"ahler potential,
$T_{\rm crit} \sim \sqrt{m_{3/2} M_\mathrm{P}}\,
(3/\beta)^{3/4}\,K''^{-1/4}$.  For realistic models we find $T_{\rm
crit} \sim 10^{11}$--$10^{12}\,$GeV, which provides an upper bound on
the temperature of the early universe. In contrast to other
cosmological constraints, this upper bound cannot be circumvented by
late--time entropy production. }

\clearpage
\section{Introduction}

Compactifications of higher--dimensional supersymmetric theories
generically contain moduli fields, which are related to approximate
vacuum degeneracy.  In many models these fields acquire masses through
condensation of fermion pairs \cite{Nilles:1982ik}, which breaks
supersymmetry. Generically, gaugino condensation models suffer from
the dilaton `run--away' problem \cite{Dine:1985he}, which can be
solved, for example, by multiple gaugino condensates
\cite{Krasnikov:jj} or non--perturbative string corrections
\cite{Shenker:1990uf,Banks:1994sg}.

Moduli play an important role in the effective low energy
theory. Their values determine geometry of the compactified space as
well as gauge and Yukawa couplings. Their masses, determined by
supersymmetry breaking, are much smaller than the compactification
scale. Hence, moduli can have important effects at low
energies. Cosmologically, they can cause the `moduli problem'
\cite{cfx83, deCarlos:1993jw}, in particular their oscillations may
dominate the energy density during nucleosynthesis, which is in
conflict with the successful BBN predictions. For an exponentially
steep dilaton potential, like the one generated by gaugino
condensation, there is also the problem that during the cosmological
evolution the dilaton ($S$) may not settle in the shallow minimum at
Re $S\sim 2$, but rather overshoot and run away to infinity
\cite{Brustein:nk}. These problems can be cured in several ways
(cf.~\cite{Dine:2000ds}).

In this paper we shall discuss a new cosmological implication of the
dilaton dynamics, the existence of a critical temperature $T_{\rm
crit}$ which represents an upper bound on allowed temperatures in the
early universe.  If exceeded, the dilaton will run to the minimum at
infinity, which corresponds to the unphysical case of vanishing gauge
couplings.  The existence of a critical temperature is a consequence
of a negative linear term in the dilaton effective potential which is
generated by finite--temperature effects in gauge theories
\cite{Buchmuller:2003is}. This shifts the dilaton field to larger
values and leads to smaller gauge couplings at high temperature. As we
shall see, this effect eventually destabilizes the dilaton, and
subsequently all moduli, at sufficiently high temperatures. In the
following we shall calculate the critical temperature $T_{\rm crit}$
beyond which the physically required minimum at Re $S\sim 2$
disappears.

There can be additional temperature--dependent contributions to the 
dilaton effective potential coming from the dilaton coupling to other scalar 
fields \cite{Huey:2000jx}. These contributions are model dependent and 
usually have a destabilizing effect on the dilaton, at least in heterotic
string models \cite{Barreiro:2000pf}. Our results for the critical
temperature can therefore be understood as conservative upper bounds on
the allowed temperatures in the early universe. 

The paper is organized as follows. In Sec.~\ref{sec:gathT} we review
the dependence of the free energy on the gauge coupling in
$\mathrm{SU}(N)$ gauge theories. As we shall see, one--loop
corrections already yield the qualitative behaviour of the full
theory.  In Sec.~\ref{sec:Tcrit} we study the dilaton potential at
finite temperature and derive the critical temperature $T_{\rm crit}$
for the most common models of dilaton stabilization.
Sec.~\ref{sec:cosmology} is then devoted to the discussion of
cosmological implications, the generality of the obtained results is
discussed in Sec.~5, and the appendix gives some details on entropy
production in dilaton decays.

\section{Gauge couplings at high temperature}
\label{sec:gathT}

The free energy of a supersymmetric $\mathrm{SU}(N_c)$ gauge theory
with $N_f$ matter multiplets in the fundamental representation reads
\begin{equation}
\mathcal{F}(g,T)\,=\,- {\pi^2 T^4 \over 24}\,
\Bigl\{\alpha_0 + \alpha_2 g^2 + {\cal O}(g^3)   \Bigr\} \;,
\label{omegagT}
\end{equation}
with $g$ and $T$ being the gauge coupling and temperature,
respectively. The zeroth order coefficient, $\alpha_0 = N_c^2 + 2 N_c
N_f - 1$, counts the number of degrees of freedom, and the one--loop
coefficient $\alpha_2$ is given by (cf.~\cite{kap89})
\begin{equation}
 \alpha_2\,=\, - {3\over 8 \pi^2}\,(N_c^2 -1)(N_c +3 N_f) \;.
\end{equation}
It is very important that $\alpha_2$ is negative.  Hence, gauge
interactions increase the free energy, at least in the weak coupling
regime. Consequently, if the gauge coupling is given by the
expectation value of some scalar field (dilaton) and therefore is a
dynamical quantity, temperature effects will drive the system towards
weaker coupling \cite{Buchmuller:2003is}.

In reality, gauge couplings are not small, e.g., $g \simeq 1/\sqrt{2}$
at the GUT scale. Thus, higher order terms in the free energy are
relevant. These could change the qualitative behaviour of the free
energy with respect to the gauge coupling. For instance, in the case
of a pure $\mathrm{SU}(N_c)$ theory, the positive $g^3$ term overrides
the negative $g^2$ term for $N_c \geq 3$.  The knowledge of higher
order terms is therefore necessary. These can be calculated
perturbatively up to order $g^6 \ln(1/g)$, where the expansion in the
coupling breaks down due to infrared divergences \cite{kap89}. The
non--perturbative contribution can be calculated by means of lattice
gauge theory. For non--supersymmetric gauge theories with matter in
the fundamental representation the free energy has been calculated up
to $g^6 \ln(1/g)$ \cite{Kajantie:2002wa}. Comparison with numerical
lattice QCD results shows that already the $g^2$ term has the correct
qualitative behaviour, i.e., gauge interactions indeed increase the
free energy.  Furthermore, if terms up to order $g^5$ are taken into
account, perturbation theory and lattice results are quantitatively
consistent, even for couplings $g = {\cal O}(1)$
\cite{Kajantie:2002wa}.

\begin{figure}[t]
\centerline{
 \subfigure[$N_f=0$.\label{fig:BNc}]{%
        \CenterObject{\includegraphics[scale=0.85]{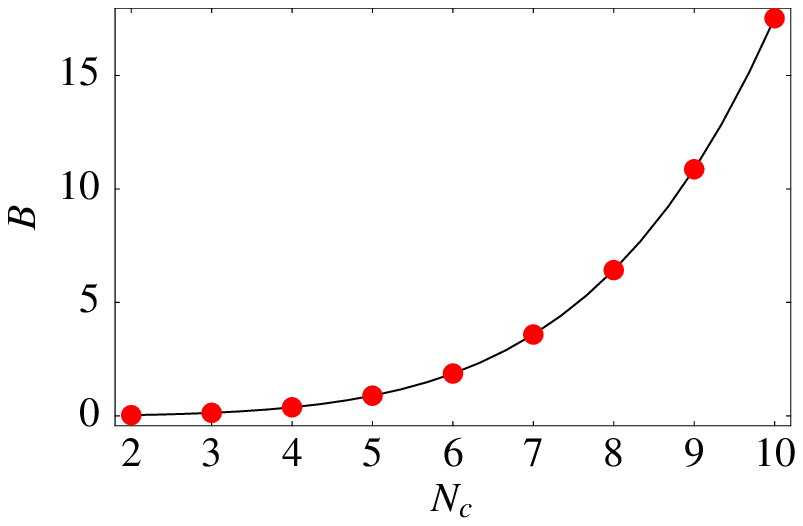}}}
 \hfil
 \subfigure[$N_c=10$.\label{fig:BNf}]{%
        \CenterObject{\includegraphics[scale=0.85]{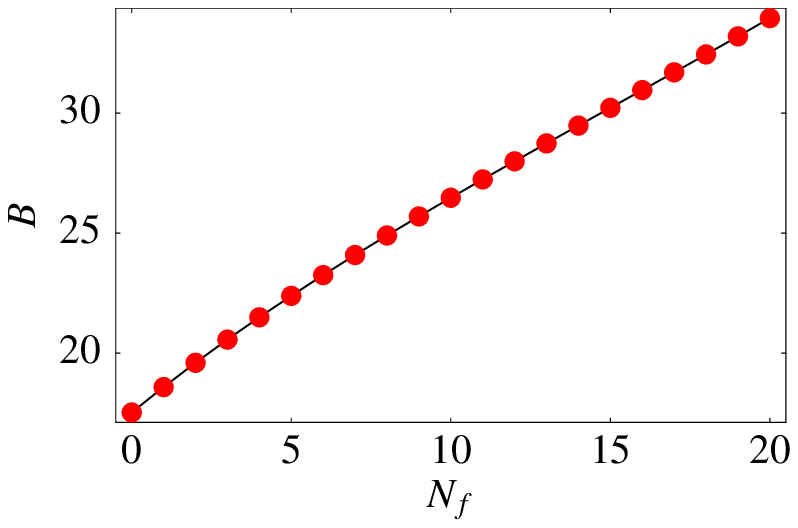}}}
}
\caption{The coefficient $B$ (cf.\ Eq.~\eqref{omega}) for
$\mathrm{SU}(N_c)$ gauge theory with $N_f$ flavours; $g_0 =
1/\sqrt{2}$.  }
\label{fig:B}
\end{figure}

To demonstrate this behaviour, we consider the free energy of a
non--supersymmetric gauge theory as a function of $N_c$ and $N_f$
using the results of Ref.\cite{Kajantie:2002wa} and earlier work
\cite{Arnold:ps}.  As discussed, it is sufficient to truncate the
perturbative expansion at order $g^5$.  We will be interested in the
free energy in the vicinity of a fixed coupling $g_0$,
\begin{eqnarray}
 g & = &  g_0 + \delta g \;,\nonumber\\
{\mathcal{F} (g,T)\over T^4} 
&=& A(g_0) + B(g_0)\, \delta g + {\cal O}(\delta g^2)\;. 
\label{omega}
\end{eqnarray}
For our purposes, it is sufficient to keep the dominant linear term
${\cal O}(\delta g)$ and neglect higher order contributions ${\cal
O}(\delta g^2)$, which have the same sign.  Fig.~\ref{fig:BNc}
displays the coefficient $B$ as a function of $N_c$ with
$N_f=0$. Analogously, Fig.~\ref{fig:BNf} shows the dependence of $B$
on the number of matter multiplets $N_f$ with $N_c =10$.  Obviously,
$B$ is positive and increases with the number of colours and
flavours. This behaviour has to be the same for all non--Abelian gauge
groups. The coefficient $B$ will be even larger in supersymmetric
theories due to gauginos and scalars.

\section{Dilaton potential at finite temperature}\label{sec:Tcrit}

\begin{figure}[t]
\centerline{\includegraphics[scale=0.9]{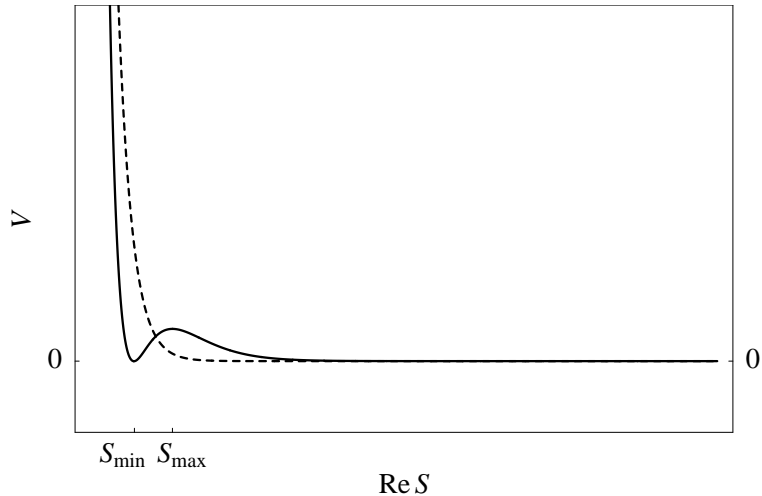}}
\caption{Typical potential for dilaton stabilization (solid curve).
A minimum at $S=S_\mathrm{min}\simeq2$ is separated from the other minimum at
$S\to \infty$ by a finite barrier. For illustration, we also plot a typical
run--away potential (dashed curve).}
\label{fig:DilatonStabilization}
\end{figure}

In this section, we discuss how finite temperature effects modify the
dilaton effective potential.  This discussion applies to many string
compactifications although details are model dependent.  The major
feature of the following analysis is that the dilaton potential has a
minimum at $\re S\sim 2$ which is separated from another minimum at
$\re S\rightarrow \infty$ by a finite barrier (see
Fig.~\ref{fig:DilatonStabilization}).  This is a rather generic
situation.

It is well known that gaugino condensation models generically suffer
from the dilaton `run--away' problem. That is, the minimum of the
supergravity scalar potential is at $S\rightarrow \infty$, i.e., zero
gauge coupling. The two most popular ways to rectify this problem in
the framework of the heterotic string use multiple gaugino condensates
\cite{Krasnikov:jj,deCarlos:1992da} and non--perturbative corrections
to the K\"ahler potential \cite{Casas:1996zi,Binetruy:1996nx}.  These
mechanisms produce a local minimum at $\re S\sim 2$.  As finite
temperature effects due to thermalized gauge and matter fields drive
the dilaton towards weaker coupling, this minimum can turn into a
saddle point, in which case the dilaton would again run away.  This
puts a constraint on the allowed temperatures in the early universe.

If the hidden sector is thermalized (cf.~\cite{Dine:2000ds}), such
constraints are meaningful as long as the temperature is below the
gaugino condensation scale, $\Lambda \sim
10^{13}$--$10^{14}\,\mathrm{GeV}$. Otherwise, by analogy with QCD, it
is expected that the gaugino condensate evaporates and the dilaton
potential vanishes.
  
The critical temperature is obtained as follows. The stabilization
mechanisms generate a local minimum of the dilaton potential at $\re
S\sim 2$, immediately followed by a local maximum, with a separation
$\delta \re S = \mathcal{O}(10^{-2})$. Beyond this local maximum, the
potential monotonously decreases to the other minimum at $\re
S\rightarrow\infty$.  Since the dilaton interaction rate $\Gamma_S
\sim T^3/M_{\rm P}^2 $ is much smaller than the Hubble parameter, the
dilaton field is not in thermal equilibrium.  It plays the role of a
background field for particles with gauge interactions since its value
determines the gauge coupling,
\begin{equation}\label{gdil}
\re S\,=\,{1\over g^2}\;.
\end{equation}
As a consequence, the complete effective potential of the dilaton field
is the sum of the zero--temperature potential $V$ and the free energy 
$\mathcal{F}$ of particles with gauge interactions,
\begin{equation}
V_T(\re S)\,=\,V(\re S) + \mathcal{F}(g=1/\sqrt{\re S},T)\;.
\end{equation}
As the temperature increases, the local minimum and maximum of $V_T$
merge into a saddle point at $\re S_{\rm crit}$. This defines the
critical temperature $T_{\rm crit}$. $\re S_{\rm crit}$ and $T_{\rm
crit}$ are determined by the two equations\footnote{In the case of
more than one solution, the maximal $T_{\rm crit}$ is the critical
temperature.}
\begin{eqnarray}
V'(\re S_{\rm crit}) + 
\mathcal{F}'(1/\sqrt{\re S_{\rm crit}},T_{\rm crit}) &=& 0 \;,\label{first}\\
V''(\re S_{\rm crit}) + 
\mathcal{F}''(1/\sqrt{\re S_{\rm crit}},T_{\rm crit}) &=& 0 \;,\label{second}
\end{eqnarray}
where `prime' denotes differentiation with respect to $\re S$.

We are only interested in the local behaviour of the potential around 
$\re S_{\rm min}\simeq 2$, where we can expand the free energy 
$\mathcal{F}(g,T)$ as in Eq.~\eqref{omega} with 
\begin{equation}
\delta g\,=\, -{\delta \re S\over 2 (\re S_{\rm min})^{3/2}} \;.
\end{equation}
This produces a linear term in $\re S$ with a 
negative slope proportional to the fourth power of the temperature,
\begin{equation}\label{linear}
\mathcal{F}(g=1/\sqrt{\re S},T)\, =\, A\ T^4  
- \delta \re S\, {1\over \xi}\, T^4  + {\cal O}((\delta \re S)^2)\;,
\end{equation}
where
\begin{equation}
\xi^{-1}\,=\,{B\over 2(\re S_{\rm min})^{3/2}} \;.
\end{equation}  
Note, that validity of the linear approximation is based on the
relation (\ref{gdil}) between the gauge coupling and the dilaton
field. In case of an arbitrary function $g = g(\re S)$ it does not
necessarily hold.

In the linear approximation the equations for the critical 
value of the dilaton field
and the critical temperature become (cf.~(\ref{first}), (\ref{second}),
(\ref{linear})),
\begin{eqnarray}
V''(\re S_{\rm crit}) &=& 0 \;,\label{scrit}\\
T_{\rm crit} &=& \left(\xi\, V'\Bigl\vert_{\re S_{\rm crit}} \right)^{1/4} \;.
\label{t}
\end{eqnarray}
At $S_{\rm crit}$, which lies between $S_{\rm min}$ and $S_{\rm max}$, the
slope of the zero--temperature dilaton potential is maximal. It is compensated
by the negative slope of the free energy at the critical temperature
$T_{\rm crit}$. For temperatures above $T_{\rm crit}$ the dilaton is driven
to the minimum at infinity where the gauge coupling vanishes.

We can now proceed to calculating the critical temperature in
racetrack and K\"ahler stabilization models. In what follows, we will
assume zero vacuum energy, which can be arranged by adding a constant
to the scalar potential. The hidden sector often contains non--simple
gauge groups, e.g. in the case of nontrivial Wilson lines. Then
gaugino condensation can occur in each of the simple factors
\cite{Krasnikov:jj}. Given the right gauge groups and matter content,
the resulting superpotential can lead to dilaton stabilization at the
realistic value of $S$ \cite{deCarlos:1992da}.  For simplicity, we
shall restrict ourselves to the case of two gaugino condensates.

The starting point is the superpotential of gaugino condensation\footnote{For 
simplicity, we neglect the Green--Schwarz 
term which would be an unnecessary complication in our analysis. 
},
\begin{equation}
W(S,\mathcal{T})\,=\,\eta(\mathcal{T})^{-6}\,\Omega(S)\;,
\end{equation}
where $\eta$ is the Dedekind $\eta$--function and 
\begin{equation}
 \Omega(S)\,=\,
 {d}_1\,\exp\left(-\frac{3  S}{2{\beta}_1}\right)
 +{d}_2\,\exp\left(-\frac{3 S}{2{\beta}_2}\right)\;.
\end{equation}
$\mathcal{T}$ is the overall T--modulus parametrizing the size of the
compactified dimensions. We assume that condensates form for two
groups, $\mathrm{SU}(N_1)$ and $\mathrm{SU}(N_2)$, with $M_1$ and $M_2$
matter multiplets in the fundamental and anti--fundamental
representations.  The parameters $d_i$ and the $\beta$--functions
$\beta_i$ are then given by ($i=1,2$),
\begin{eqnarray}
 {\beta}_i
 & = &
 \frac{3N_i-M_i}{16\pi^2}\;,\\
 {d}_i
 & = & 
 \left(\tfrac{1}{3}M_i-N_i\right)\,
 \left(32\pi^2\,e\right)^{3(M_i-N_i)/(3N_i-M_i)}\,
 \left(\tfrac{1}{3}M_i\right)^{M_i/(3N_i-M_i)}\;.
\end{eqnarray}

Together with the K\"ahler potential
\begin{equation}
\mathcal{K}\,=\,
K(S + \bar S) - 3 \ln{(\mathcal{T} + \overline{\mathcal{T}})}\;,
\end{equation}
the superpotential for gaugino condensation yields the scalar 
potential \cite{deCarlos:1992da},
\begin{equation}\label{scapot1}
 V\,=\,
 \frac{|\eta(\mathcal{T})|^{-12}}{(2 \re \mathcal{T})^3}\,e^K\,
 \left\{{1\over K_{S \bar S}}
  |\Omega_S + K_S \Omega|^2
  +\left(\frac{3(\re \mathcal{T})^2}{\pi^2}|\widehat{G}_2|^2-3\right)
  |\Omega|^2
 \right\}\;,
\end{equation}
where subscripts denote differentiation with respect to the specified 
arguments, and the function $\widehat{G}_2$ is defined via the Dedekind 
$\eta$--function as
\begin{equation}
 \widehat{G}_2\,=\,
 -\left(\frac{\pi}{\re\mathcal{T}}+
        4\pi\,\frac{\eta'(\mathcal{T})}{\eta(\mathcal{T})}\right)\;.
\end{equation}

It is well known that the T--modulus settles at a value $\mathcal{T}
\sim 1$ in Planck units, independently of the condensing gauge groups
\cite{deCarlos:1992da}.  Further, in the case of two condensates,
minimization in $\im S$ simply leads to opposite signs for the two
condensates in $\Omega$. From Eq.~(\ref{scapot1}) we then obtain a
scalar potential which only depends on $x \equiv \re S$, the real part of
the dilaton field,
\begin{equation}\label{scapot2}
 V(x)\,=\,
 a\, e^K\,\left({4\over K''}
  \left(\Omega' + {1\over 2} K' \Omega\right)^2 - b\ \Omega^2 \right)\;,
\end{equation}
where $a \simeq b \simeq 3$ and
\begin{equation}
 \Omega(x)\,=\,
 {d}_1\,\exp\left(-\frac{3  x}{2{\beta}_1}\right)
 -{d}_2\,\exp\left(-\frac{3 x}{2{\beta}_2}\right)\;.
\end{equation}
The dilaton is stabilized at a point
$x_{\rm min}$ where the first derivative of the potential,
\begin{eqnarray}
V'\, &=&\,
2 a\, e^K\, \left(\Omega' + {1\over 2} K' \Omega\right)
\left\{ \left(\Omega' + {1\over 2} K' \Omega\right) 
\left(4 {K'\over K''} - 2 {K'''\over K''^2}\right)\right. \nonumber\\
&& \left.\hspace{4cm} {}
+ {4\over K''}\Omega'' - \left({K'^2\over K''} + b-2\right)\ \Omega \right\}\;,
\end{eqnarray}
vanishes, and the dilaton mass term is positive,
\begin{equation}
m_S^2\,=\,2 \left.{V''\over K''}\right|_{x_{\rm min}}\,>\,0\;.
\end{equation}
In the following we shall determine the critical temperature for two
models of dilaton stabilization. The scales of dilaton mass and
critical temperature are set by the gravitino mass,
\begin{equation}
 m_{3/2}^2
 \,=\,
 e^{\mathcal{K}} \vert W \vert^2\,\Bigl\vert_{x_{\rm min}}
 \,=\,
 a\, e^{K} \vert \Omega \vert^2\,\Bigl\vert_{x_{\rm min}}\;,
 \label{m32}
\end{equation}
and the scale of supersymmetry breaking, 
$M_{\rm SUSY} = \sqrt{m_{3/2}}$, measured in Planck units.

\subsection{Critical temperature for racetrack models}

Consider first the case with the standard K\"ahler potential,
\begin{equation}
 K(S + \bar S)\, =\, - \ln{(S + \bar S)}\;,
\end{equation}
and two gaugino condensates, the so-called `racetrack models'.
The first derivative of the scalar potential (\ref{scapot2}) then becomes
\begin{equation}\label{d1pot}
 V'\, =\,
 2 a\, e^K\, \left(\Omega' + {1\over 2} K' \Omega\right)
 \left({4\over K''}\Omega'' - (b-1)\ \Omega \right)\;.
\end{equation}
It has been shown \cite{deCarlos:1992da} that the local minimum is determined 
by the vanishing of the first factor,
$(2\,x\,\Omega'(x)-\Omega(x))|_{x_{\rm min}}=0$.

\begin{figure}[t]
\centerline{
 \subfigure[$T=0$.]{%
\CenterObject{\includegraphics[scale=0.85]{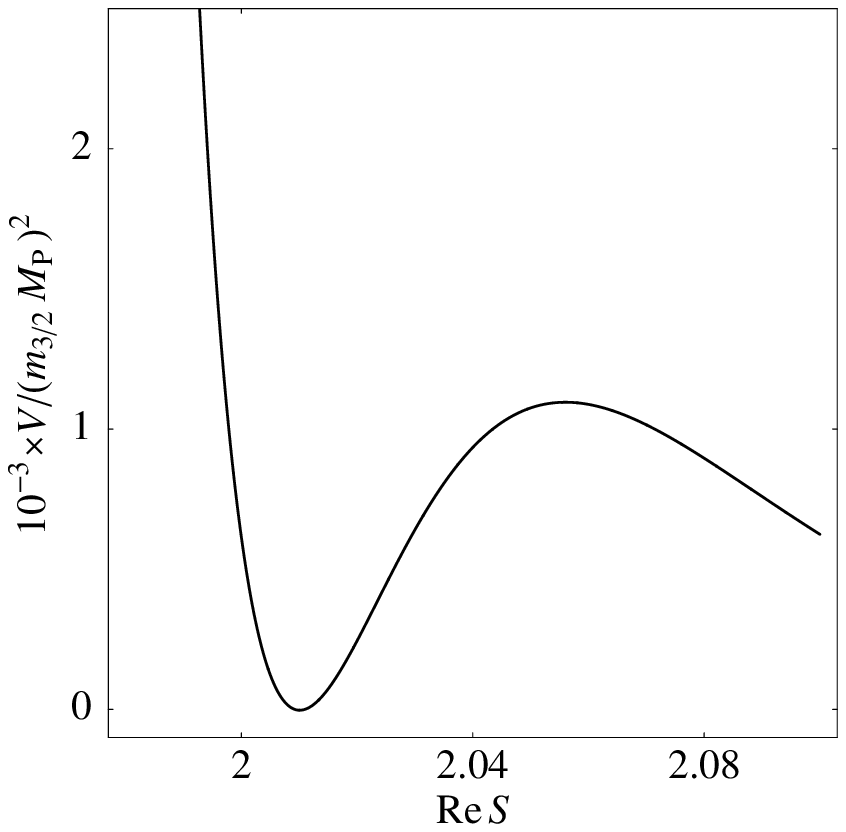}}}
 \hfil
 \subfigure[$T=T_\mathrm{crit}$.]{%
\CenterObject{\includegraphics[scale=0.85]{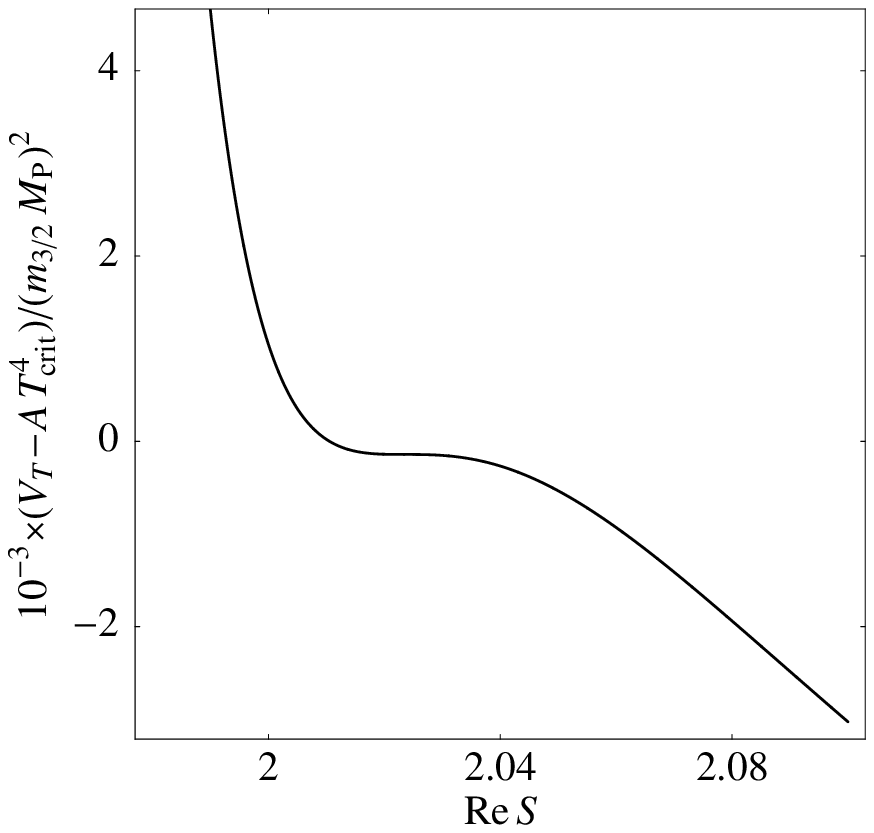}}}
}
\caption{Dilaton potential for $(N_1,N_2)=(7,8)$ and
$(M_1,M_2)=(8,15)$.  (a): $T=0$, (b): $T=T_\mathrm{crit}$. In (b) the
dilaton independent term $A\, T_\mathrm{crit}^4$ has been subtracted
(cf.\ Eq.~\eqref{linear}). }
\label{fig:DilatonPotential}
\end{figure}

We now have to evaluate (\ref{d1pot}) at the point of zero curvature,
$V''=0$.  Differentiation by $x$ brings down a power of $3/(2 \beta)
\gg 1$.  Away from the extrema, where cancellations occur, we
therefore have the following hierarchy,
\begin{equation}
 \vert \Omega \vert
 \,\ll\, 
 \vert \Omega' \vert
 \,\ll\, 
 \vert \Omega'' \vert 
 \,\ll\,
 \vert \Omega''' \vert \;.
\end{equation}
This implies for the first and second derivative of the potential,
\begin{eqnarray}
 V' &\simeq& 2a\, e^K\,{4\over K''}\,\Omega'\,\Omega''\;,\\
 V'' &\simeq& 2a\, e^K\,{4\over K''}\,
 \left(\Omega''^2 + \Omega'\,\Omega'''\right)\;.
\end{eqnarray}
For the slope of the potential at the critical point one then obtains the
convenient expression
\begin{equation}
 V'\Bigl\vert_{x_{\rm crit}} \,\simeq\,
 - 2 a\,e^K\,{4\over K''} {(\Omega')^2\,\Omega''' \over \Omega''} \;.
\end{equation}
For $x_{\rm min} < x < x_{\rm max}$ one has
\begin{equation}
 \Omega'
 \,\sim\, 
 -{3\over 2\beta_{\rm max}}\,\Omega\;, \quad
 \Omega'''
 \,\sim\,
 -{3\over 2\beta_{\rm min}}\,\Omega''\;, 
\end{equation}
where $\beta_{\rm max}$ ($\beta_{\rm min}$) is the larger (smaller) of the
two $\beta$--functions. This yields for the slope of the potential
\begin{equation}
 V'\Bigl\vert_{x_{\rm crit}} 
 \,\sim\,
 2 a\,e^K\,{4\over K''} \left({3\over 2\beta_{\rm max}}\right)^2 
 \left({3\over 2\beta_{\rm min}}\right) \Omega^2 \;.
\end{equation}
Since $\Omega$ does not vary significantly 
between $x_{\rm min}$ and $x_{\rm crit}$,
one finally obtains (cf.~(\ref{m32})),
\begin{equation}\label{vprt}
 V'\Bigl\vert_{x_{\rm crit}} 
 \,\sim\,
 {1\over K''}\,\left({3\over \beta_{\rm max}}\right)^2\, 
 \left({3\over \beta_{\rm min}}\right)\, m_{3/2}^2 \;.
\end{equation}

Using Eq.~(\ref{t}) we can now write down the critical temperature. Note
that in racetrack models $\beta_{\rm min}$ and $\beta_{\rm max}$ are
usually very similar. Introducing
$\beta = (\beta_{\rm min} \beta_{\rm max}^2)^{1/3}$, one obtains
\begin{equation}\label{tcrit1}
 T_{\rm crit}
 \,\sim\,
 \sqrt{m_{3/2}}\, \left(3\over\beta\right)^{3/4}
 \left({\xi\over K''}\right)^{1/4}\;.
\end{equation}
We have determined $T_{\rm crit}$ also numerically. The result agrees
with Eq.~(\ref{tcrit1}) within a factor $\sim 2$.  The factor
$\sqrt{m_{3/2}}$ appears since the scale of the scalar potential is
set by $m_{3/2}^2$. The $\beta$--function factor corrects for the
steepness of the scalar potential, whereas $(\xi/ K'')^{1/4} = {\cal
O}(1)$. With $m_{3/2}\sim 100\,$GeV, $\beta \sim 0.1$ and
$M_\mathrm{P}=2.4\times 10^{18}\,\mathrm{GeV}$, one obtains
\begin{equation}
 T_{\rm crit}
 \,\sim\,
 10^{11}\,{\rm GeV}\;,
\end{equation}
as a typical value of the critical temperature.

A straightforward calculation yields for the dilaton mass
\begin{equation}\label{md1}
 m_S
 \,\simeq\, 
 {9 \over \beta_1 \beta_2}{1 \over K''} \, m_{3/2}\;.
\end{equation}
As a result, the dilaton mass is much larger than the gravitino mass
and lies in the range of hundreds of TeV. This fact will be important
for us later when we discuss the $S$--modulus problem.

\subsection{Critical temperature for K\"ahler stabilization}

As a second example we consider dilaton stabilization through
non--perturbative  corrections  to the K\"ahler potential. 
In this case a single gaugino condensate is
sufficient \cite{Casas:1996zi,Binetruy:1996nx}. Like instanton
contributions, such corrections are expected to vanish in the limit of 
zero coupling and also to all orders of perturbative expansion. 
A common parametrization of the non--perturbative  corrections reads
\begin{eqnarray}
 e^K & = & e^{K_0}+ e^{K_{\rm np}} \;, \nonumber\\
 e^{K_{\rm np}} & = & c\, x^{p/2}\, e^{-q \sqrt{x}}\;,
\end{eqnarray}
with $K_0=-\ln(2x)$, $x=\re S$, and parameters subject to $K'' > 0$ and 
$p,q > 0$. For a single gaugino condensate, one has
\begin{equation}
 \Omega\,=\,d\,\exp{\left(-{3x\over 2\beta}\right)}\;,
\end{equation}
where $3/(2\beta)=8\pi^2/N$ and $d=-N/(32 \pi^2 e)$ for a condensing
$\mathrm{SU}(N)$ group with no matter. Note that the scalar potential
is independent of $\im S$.

\begin{figure}[t]
\centerline{
 \subfigure[$T=0$.]{%
\CenterObject{\includegraphics[scale=0.85]{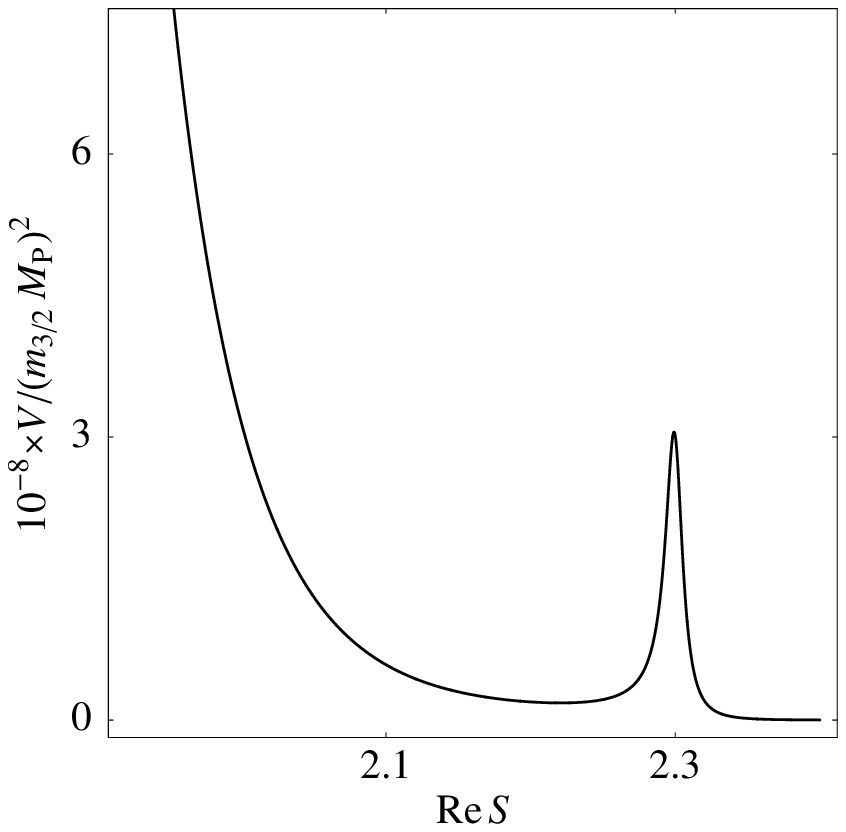}}}
 \hfil
 \subfigure[$T=T_\mathrm{crit}$.]{%
\CenterObject{\includegraphics[scale=0.85]{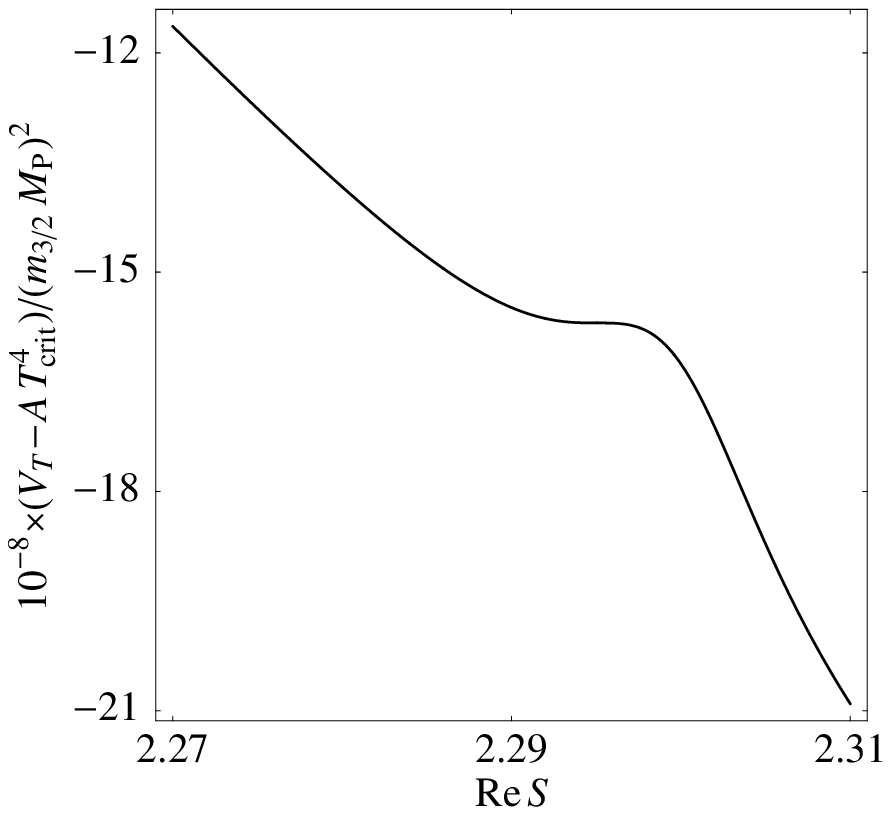}}}
}
\caption{Dilaton potential for K\"ahler stabilization. $c=5.7391$,
  $p=1.1$, $q=1$, and $N=6$ \cite{Barreiro:1997rp}.  (a): $T=0$, (b):
  $T=T_\mathrm{crit}$.  In (b) the dilaton independent term $A\,
  T_\mathrm{crit}^4$ has been subtracted (cf.\ Eq.~\eqref{linear}). }
\label{fig:DilatonPotentialKahler}
\end{figure}

The scalar potential and its derivative are given by the simple expressions
\begin{eqnarray}\label{scapotks}
 V(x)\,&=&\,
 a\, e^K\,\Omega^2\, \left({1\over K''}
  \left(K' - {3\over \beta}\right)^2 - b \right)\;, \\
V'(x) &=& a\, e^K\,\Omega^2\, \left(K'- {3\over \beta}\right)
\left( {1\over K''} \left(K'- {3\over \beta}\right)^2 \right. \nonumber\\
&& \left.\hspace{3.5cm}
{}- {K''' \over (K'')^2}\left(K'- {3\over \beta}\right) - b+2\right)\;.
\label{kahlerVp}
\end{eqnarray}
It has been shown [\citen{Barreiro:1997rp},\citen{Casas:1996zi}] that
realistic minima are associated with the singularity at $K''=0$. That
is, by tuning the parameters $c,p,q$ it is possible to adjust $K''=0$
at some value $x$ where the potential then blows up. By perturbing the
parameters slightly, one obtains a finite potential with positive but
small $K''$, and the singularity smoothed out into a finite bump.  The
bump is located approximately at the point of minimal $K''$, and the
local minimum of the potential at $x\sim 2$ lies very close to it,
with a separation $\delta x = \mathcal{O}(10^{-2})$.

For realistic cases, $K'(x\sim 2) \ll 3/\beta$, and the 
extrema of the potential around $x\sim 2$ are associated
with the zeros of the last bracket in Eq.~(\ref{kahlerVp}).
As explained above, in practice $K''$ is a very small parameter
such that one  can expand in powers of $K''$. Then,  the extrema
appear due to cancellations between the two `singular' terms and
we have the approximate relation
\begin{equation}\label{kder}
 {K'''\over K''}\,\simeq\, -{3\over \beta}  \;.
\end{equation}
Due to the spiky shape of the potential, the point of vanishing
curvature, $V''=0$, is very close to the local maximum.
On the other hand, the cancellations between the $1/K''$ and 
$1/(K'')^2$ terms in  Eq.~(\ref{kahlerVp}) are not precise
at this point and one can approximate their sum by the
larger term. Using the fact that $K$ and $\Omega$ do not vary significantly
between $x_{\rm min}$ and $x_{\rm crit}$, one obtains from Eqs.~(\ref{m32})
and (\ref{kder}),
\begin{equation}
 V'\Bigl\vert_{x_{\rm crit}}
 \,\sim\,
 a\,e^K\,{1\over K''} \left({3\over \beta}\right)^3 \Omega^2 
 \,\sim\, 
 {1\over K''} \left({3\over \beta}\right)^3\,m_{3/2}^2 \;, 
\end{equation}
where $K''$ is evaluated at the local maximum $x_{\rm max}$. 
Note that this result is identical to Eq.~(\ref{vprt}) which we have obtained
for racetrack models. However, for these models 
$1/(K'')^{1/4}=\sqrt{x}={\cal O}(1)$, whereas now $K''$ is a very small,
but otherwise essentially free parameter.

Using Eq.~(\ref{t}) we find the same expression for the critical temperature
as in racetrack models,
\begin{equation}\label{tcritks}
 T_{\rm crit}
 \,\sim\,
 \sqrt{m_{3/2}}\, \left(3\over\beta\right)^{3/4}  
 \left({\xi\over K''}\right)^{1/4} \;.
\end{equation}
Since $K''$ is small in realistic models, the upper bound on allowed
temperatures relaxes compared to racetrack models.  As before,
Eq.~(\ref{tcritks}) agrees within a factor $\sim 2$ with numerical
results.  A typical value of the critical temperature is obtained for
$m_{3/2}\sim 100$ GeV, $\beta \sim 0.1$ and $K''\sim 10^{-4}$,
\begin{equation}
 T_{\rm crit}
 \,\sim\, 
 10^{12}\, {\rm GeV} \;.
\end{equation}
For the dilaton mass one obtains
\begin{equation}
 m_S\,\sim\, \left(\frac{3}{\beta}\right)^2\,\frac{1}{K''}\,m_{3/2}\;.
\end{equation}
Again, we find that the dilaton is much heavier than the gravitino.

\section{Implications for cosmology}
\label{sec:cosmology}

As we have seen in the previous section, the dilaton gets destabilized
at high temperature.  The maximal allowed temperature is given by
$T_\mathrm{crit}\sim 10^{11}\mathrm{-}10^{12}\,\mathrm{GeV}$.  In
this section, we study implications of this bound for cosmology.

Most importantly, $T_\mathrm{crit}$ represents a model independent
upper bound on the temperature of the early universe,
\begin{equation}
 T \,<\, T_\mathrm{crit}\;.
\end{equation} 
This bound applies to a large class of theories, with weakly coupled
heterotic string models being the most prominent representatives.  It
is worth emphasizing that the dilaton destabilization effect is
qualitatively different from the gravitino \cite{gravitino} or moduli
problems \cite{cfx83,deCarlos:1993jw} in that it cannot be
circumvented by invoking other effects in late--time cosmology such as
additional entropy production.  Once the dilaton goes over the
barrier, it cannot come back.

The present bound applies to any radiation dominated era in the early
universe, even if additional inflationary phases occur
afterwards. Therefore, $T_\mathrm{crit}$ not only provides an upper
bound on the reheating temperature $T_R$ of the last inflation, but
also can be regarded as an absolute upper bound on the temperature of
the radiation dominated era in the history of the universe.

\subsection{$\boldsymbol{S}$--modulus problem and thermal leptogenesis}
\label{subSec-Smoduli}

In addition to the bound discussed above, one can have further, more
model dependent, constraints on temperatures occurring at various
stages of the evolution of the universe.  In this subsection, we
discuss one of them, related to the $S$--modulus problem.

Even if the reheating temperature does not exceed the critical one,
thermal effects shift the minimum of the dilaton potential.  Due to
this shift, $S$ starts coherent oscillations after reheating.  Since
the energy density stored in the oscillations behaves like
non--relativistic matter, $\rho_\mathrm{osc}\propto R^{-3}$, with $R$
being the scale factor, it grows relative to the energy density of the
thermal bath, $\rho_\mathrm{rad}\propto R^{-4}$, until $S$ decays.
Its lifetime can be estimated as $(\Gamma_S)^{-1}\sim
M_\mathrm{P}^2/m_S^3 \simeq
0.004\,\mathrm{s}\,(m_S/100\,\mathrm{TeV})^{-3}$.  In the examples
studied in Sec.~\ref{sec:Tcrit}, $m_S\gg 10\,\mathrm{TeV}$, so that
$S$ decays before BBN.  Thus, there is no conventional moduli problem,
i.e., dilaton decays do not spoil the BBN prediction of the abundance
of light elements.

However, even for these large masses, coherent oscillations of $S$ may
affect the history of the universe via entropy production
\cite{cfx83,deCarlos:1993jw}.  Let us estimate the initial amplitude
of these oscillations.  At a given temperature $T\ll T_{\rm crit}$,
the dilaton potential around the minimum can be recast as
\begin{equation}
  V_T\,=\,\frac{1}{2}m_S^2\,\phi^2
 -\sqrt{\frac{2}{\xi^2\,K''}}T^4\,\frac{\phi}{M_\mathrm{P}}\;,
 \label{eq-VT}
\end{equation}
where $\phi=M_\mathrm{P}\sqrt{K''/2} \,\re (S-S_\mathrm{min})$. The
minimum of the potential is at
\begin{equation}\label{eq:phiT}
 \langle\phi\rangle_T\,\simeq\,
 \sqrt{\frac{2}{\xi^2\,K''}}
 \frac{T^4}{m_{S}^2\,M_\mathrm{P}}\;.
\end{equation}
Thus, at $T = T_R$, the displacement of $\phi$ from its zero
temperature minimum is estimated as $\Delta\phi|_{T_R}\sim
\langle\phi\rangle_{T_R}$.  Then, the entropy produced in dilaton
decays is (see Appendix),
\begin{equation}\label{eq:dilution}
 \Delta
 \,=\,
 \frac{s_\mathrm{after}}{s_\mathrm{before}}
 \,\sim\,
 \frac{1}{\xi^2 K''}
 \left(\frac{T_R}{10^{10}\,\mathrm{GeV}}\right)^5
 \left(\frac{10^6\,\mathrm{GeV}}{m_S}\right)^{7/2}\;.
\end{equation}
The decay occurs at temperatures of order $10\,\mathrm{MeV}$, i.e.,
after the baryon asymmetry and the dark matter abundance have been
fixed.  Thus, we see that for $T_R\gtrsim
10^{10}\,\mathrm{GeV}\,(m_S/10^6\,\mathrm{GeV})^{7/10} (\xi^2
K'')^{1/5}$, the baryon asymmetry and relic dark matter density get
significantly diluted.

For instance, successful thermal leptogenesis \cite{FY} requires
$T_R\gtrsim T_\mathrm{L}\simeq\,3\times 10^9\,\mathrm{GeV}$
\cite{Buchmuller:2004nz}.  For $T_R\gtrsim T_\mathrm{L}$, the baryon
asymmetry can be enhanced by $T_R/T_\mathrm{L}$, but later it gets
diluted by a factor $\propto T_R^5$.  Hence, there is only a narrow
temperature range where thermal leptogenesis is compatible with the
usual mechanisms of dilaton stabilization.  We note further that, in
this range of temperatures, the bound on the light neutrino masses
tightens.  For instance, $T_R<3\times 10^{10}\,\mathrm{GeV}$
implies\footnote{Here we have used Fig.~10 of
Ref.~\cite{Buchmuller:2004nz},
$m_1<\widetilde{m}_1$ \cite{Fujii:2002jw}, and $m_3^2-m_1^2\simeq
\Delta m^2_\mathrm{atm}$.}  $m_i\lesssim 0.07\,\mathrm{eV}$, which is
more stringent than the temperature--independent constraint,
$m_i\lesssim 0.1\,\mathrm{eV}$ \cite{Buchmuller:2003gz}.

Concerning dark matter, we note that in WIMP cold dark matter
scenarios, at the time of the dilaton decay the pair annihilation
processes have frozen out so that the entropy production reduces
$\Omega_\mathrm{CDM}$.\footnote{WIMP dark matter may be directly
produced by moduli decay \cite{Moroi:1999zb}.}  This effect could be
welcome in parameter regions where otherwise WIMPs are overproduced.
Entropy production could also contribute to the solution of the
gravitino problem.

It is important to remember that the T--moduli problem
remains. Thermal effects shift all moduli from their zero temperature
minima, thereby inducing their late coherent oscillations.  Unlike the
dilaton, other moduli typically have masses of order $m_{3/2}$ and
thus tend to spoil the BBN predictions.

In summary, there exists a range of reheating temperatures,
$10^{-2}T_\mathrm{crit}\lesssim T_R\lesssim T_\mathrm{crit}$, which are
cosmologically acceptable, but for which the history of the universe
is considerably altered, in particular via significant entropy
production at late times.

\subsection{Further constraints on inflation models}

\begin{figure}[t!]
\centerline{
 \CenterObject{\includegraphics[scale=1.1]{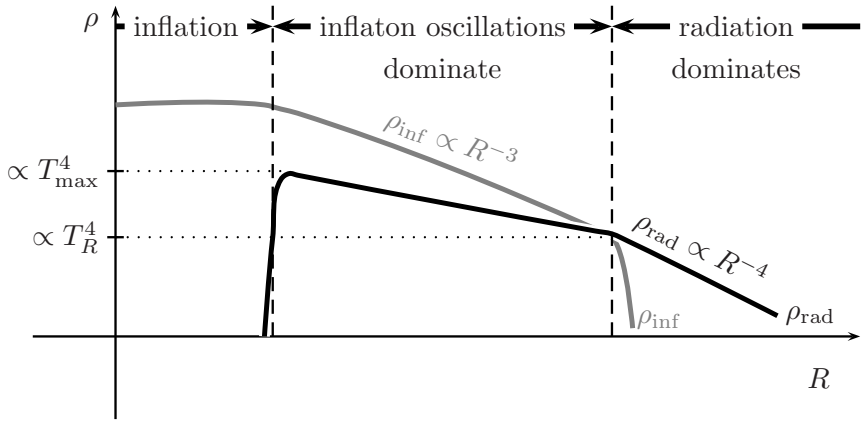}}
}
\caption{Three epochs in inflationary models: inflation, inflaton oscillation
domination and radiation domination \cite{Kolb-Turner}.}
\label{fig:FromInflationToReheating}
\end{figure}

In this subsection we discuss some implications of the thermal effects
at earlier times, before the reheating process completes.  There are
three important stages in the inflationary scenario: inflation, the
inflaton--oscillation epoch, and the radiation dominated epoch
(see Fig.~\ref{fig:FromInflationToReheating}).

During inflation, the energy density of the universe is dominated by
the potential energy of the inflaton $\chi$. After the end of
inflation, inflaton starts its coherent oscillations. The energy
density of the universe is still dominated by the inflaton $\chi$,
until the reheating process completes and radiation energy takes over
with temperature $T=T_R$.  The nonzero energy density of the inflaton
induces additional SUSY breaking effects \cite{DineRandallThomas}.  
Hence, one may expect that the
dilaton potential is also affected by the finite energy of the
inflaton $\chi$ during these $\chi$--dominated eras.

Further, in the $\chi$--oscillation era there is radiation with
temperature $T\simeq (T_R^2 M_\mathrm{P} H)^{1/4}$ \cite{Kolb-Turner},
where $H$ is the Hubble parameter. Although its energy density is
small compared to that of inflaton (see
Fig.~\ref{fig:FromInflationToReheating}), it affects the dilaton
potential as we have discussed in Sec.~\ref{sec:Tcrit}.  Since the
maximum temperature $T_\mathrm{max}$ in the $\chi$--oscillation era is
generically higher than the reheating temperature $T_R$, one expects
stronger constraints from $T_\mathrm{max}<T_\mathrm{crit} $.

Whether it is radiation or inflaton that affects the dilaton potential
more, depends on the coupling between dilaton and inflaton.  As this
is model dependent, below we consider the three possible cases:

{\bf (i) destabilizing dilaton--inflaton coupling}. 
The inflaton--dilaton coupling   
drives the dilaton to larger values
and may let it run away to infinity.
This puts severe constraints on the inflation model.
Some
models can be excluded independently of the reheating temperature.

{\bf (ii) stabilizing dilaton--inflaton coupling}.  The inflaton
effects move the dilaton to smaller values.  In this case, the
previously obtained bound on the reheating temperature $T_R <
T_\mathrm{crit}$ provides the most stringent constraint. Note that the
shift of the dilaton may cause a large initial amplitude of its
oscillation, which can result in a late--time entropy production as
discussed in Sec.~\ref{subSec-Smoduli}.

\begin{figure}[t!]
\centerline{
\CenterObject{\includegraphics[scale=0.9]{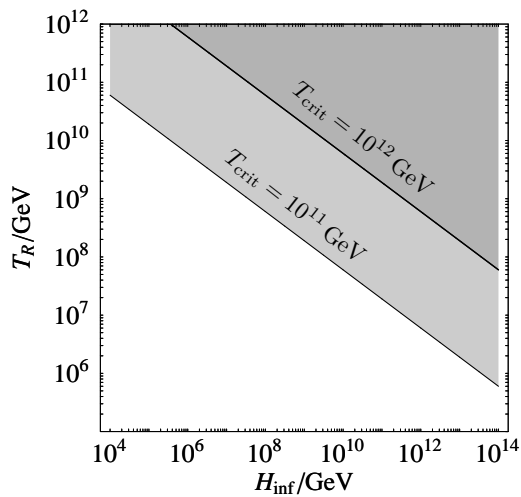}}
}
\caption{Upper bounds on the reheating temperature for
 $T_\mathrm{crit} = 10^{11}\,\mathrm{GeV}$ and $10^{12}\,\mathrm{GeV}$
assuming a small inflaton--dilaton coupling
(case (iii)). 
 The darker region is excluded for
 $T_\mathrm{crit}=10^{12}\,\mathrm{GeV}$; for 
 $T_\mathrm{crit}=10^{11}\,\mathrm{GeV}$ the lighter region is  
excluded as well.}
\label{fig:Tmax}
\end{figure}

{\bf (iii) negligible dilaton--inflaton coupling}.  In this case, the
effect of radiation during the $\chi$--oscillation era (preheating
epoch) is dominant.  The maximal radiation temperature can be
expressed in terms of the reheating temperature \cite{Kolb-Turner},
\begin{equation}
  T_\mathrm{max}\,\simeq\, (T_R^2 M_\mathrm{P} H_\mathrm{inf})^{1/4}\;,
\end{equation}
where $H_\mathrm{inf}$ is the Hubble expansion rate during inflation. 
$T_\mathrm{max}$ must be below the critical temperature,
or the dilaton will run away to weak coupling.
This constraint translates into a bound on the reheating
temperature depending on $H_\mathrm{inf}$,
\begin{equation}
  T_R \,\lesssim\, 
  \left(\frac{T_\mathrm{crit}^4}{M_\mathrm{P}H_\mathrm{inf}}\right)^{1/2}
  \,\simeq\,6\times 10^7\,\mathrm{GeV}
  \left(\frac{T_\mathrm{crit}}{10^{11}\,\mathrm{GeV}}\right)^2
  \left(\frac{10^{10}\,\mathrm{GeV}}{H_\mathrm{inf}}\right)^{1/2}\;,
\end{equation}
as shown in Fig.~\ref{fig:Tmax}.
The upper bound on $T_R$ now becomes much severer. For instance,
taking $T_\mathrm{crit}\simeq 10^{11}\,\mathrm{GeV}$ and typical
values of $H_\mathrm{inf}$ in some inflation models~\footnote{In
curvaton scenarios \cite{curvaton} the values of $H_\mathrm{inf}$ are
much less constrained.}  (cf.~\cite{Asaka:1999jb}), we obtain the
following bounds:\\
\begin{tabular}{lll}
 chaotic inflation: & 
 $H_\mathrm{inf}\simeq 10^{13}\,\mathrm{GeV}$, &
 $T_R\lesssim 10^6\,\mathrm{GeV}$,
 \\
 hybrid inflation: &
 $H_\mathrm{inf}\sim 10^8$--$10^{12}\,\mathrm{GeV}$, &
 $T_R\lesssim 10^7$--$10^9\,\mathrm{GeV}$,
 \\
 new inflation: &
 $H_\mathrm{inf}\sim 10^6$--$10^{12}\,\mathrm{GeV}$, &
 $T_R\lesssim 10^7$--$10^{10}\,\mathrm{GeV}$.
\end{tabular}
\\ These bounds apply if already during the preheating phase particles
with gauge interactions form a plasma with temperature
$T_\mathrm{max}$ and the dilaton is near the physical minimum.

\section{Conclusions}

At finite temperature the effective potential of the dilaton acquires
locally a negative linear term. As we have seen, this important fact
is established beyond perturbation theory by lattice gauge theory
results.  As a consequence, at sufficiently high temperatures the
dilaton $S$, and subsequently all other moduli fields, are
destabilized and the system is driven to the unphysical ground state
with vanishing gauge coupling.  We have calculated the corresponding
critical temperature $T_{\rm crit}$ which is larger than the scale of
supersymmetry breaking, $M_\mathrm{SUSY} = \sqrt{M_{\rm P} m_{3/2}} =
\mathcal{O}(10^{10}\,{\rm GeV})$, but significantly smaller than the
scale of gaugino condensation, $\Lambda =
[d\,\exp(-3S/(2\beta))]^{1/3} M_\mathrm{P} =
10^{13}$--$10^{14}\,\mathrm{GeV}$.  This is the main result of our
paper.

Our result is based on the well understood thermodynamics of the
observable sector. In contrast, the temperature of gaugino
condensation can place a bound on the temperature of the
early universe only under the additional assumption that the hidden
sector is thermalized.

The upper bound on the temperature in the radiation dominated phase of
the early universe, $T < T_{\rm crit} \sim 10^{11}$--$10^{12}\,$GeV,
has important cosmological implications.  In particular, it severely
constrains baryogenesis mechanisms and inflation scenarios. Models
requiring or predicting $T>T_\mathrm{crit}$ are incompatible with
dilaton stabilization. In contrast to other cosmological constraints,
this upper bound cannot be circumvented by late--time entropy
production.

We have also discussed more model dependent cosmological
constraints. Even if $T<T_\mathrm{crit}$, the $S$--modulus problem
restricts the allowed temperature of thermal leptogenesis and makes
the corresponding upper bound on light neutrino masses more
stringent. Furthermore, depending on the assumed coupling between
dilaton and inflaton, stronger bounds can apply to the reheating
temperature.

Our discussion has been based on the assumption that moduli are
stabilized by non--perturbative effects which break supersymmetry.
Thus the barrier separating the realistic vacuum from the unphysical
one with zero gauge couplings is related to supersymmetry breaking.
Recently, an interesting class of string compactifications has been
discussed where fluxes lead to moduli stabilization and supersymmetry
breaking (see, for example, \cite{gkp02,kkx03,ggx03}). Realistic,
metastable de Sitter vacua also require non--perturbative
contributions to the superpotential from instanton effects or gaugino
condensation \cite{kkx03}. It remains to be seen how much fluxes can
modify the critical temperature in realistic string compactifications.

\subsection*{Acknowledgments}
We would like to thank A.~Hebecker, T.~Kobayashi, J.~Louis and
T.~Moroi for helpful comments.

\subsection*{Appendix\quad Evolution of $\boldsymbol{\phi}$ 
and entropy production}

In the following, we derive the dilution factor,
Eq.~\eqref{eq:dilution}.  The dilaton starts coherent oscillations
soon after the radiation dominated era begins.  This is because the
effect of the temperature term in Eq.~\eqref{eq-VT} disappears very
quickly and when $|\phi-\langle\phi\rangle_T|\gtrsim
\langle\phi\rangle_T$ the potential becomes essentially quadratic.  As
can be verified numerically, the initial amplitude of subsequent
oscillations is close to the initial displacement of the dilaton from
its zero temperature minimum, $\Delta
\phi|_{T_R}\sim\langle\phi\rangle_{T_R}$.  The Hubble friction is very
small at these times, $H\ll m_S$.

The ratio of $\rho_\mathrm{osc}$ to the entropy density $s$ before the
dilaton decays is given by (cf.~Eq.~\eqref{eq:phiT}),
\begin{eqnarray}
  \left.\frac{\rho_\mathrm{osc}}{s}\right|_\mathrm{before}
  \,\sim\,
  \frac{m_S^2\langle\phi\rangle_{T_R}^2}{s(T_R)}
  \,=\,
  \frac{2T_R^5}{(2\pi^2/45)\,g_*(T_R)\,\xi^2\, K''\,m_S^2\, M_\mathrm{P}^2}\;.
\end{eqnarray}
The ratio stays constant since $\rho_\mathrm{osc}\propto s\propto R^{-3}$. 
Just after the dilaton  decays, the ratio of $\rho_\mathrm{rad}$ to $s$ is
\begin{eqnarray}
  \left.\frac{\rho_\mathrm{rad}}{s}\right|_\mathrm{after}
  \,=\,
  \frac{3}{4}T_d
  \,\simeq\,
  \frac{3}{4}\left(\frac{\pi^2}{90}\,g_*(T_d)\right)^{-1/4}
  \sqrt{\Gamma_S\, M_\mathrm{P}} 
  \;.
\end{eqnarray}
If $\rho_\mathrm{osc}/s > \rho_\mathrm{rad}/s$, the dilaton decay
causes large entropy production.  Using
$\rho_\mathrm{rad}|_\mathrm{after} \simeq
\rho_\mathrm{osc}|_\mathrm{before}$, we obtain
Eq.~\eqref{eq:dilution}.  Note that there are large numerical
uncertainties in this expression due to the dependence on initial
conditions.  In extreme scenarios, $\Delta$ can be close to one.
However, the resulting uncertainty in $T_R$ is usually rather small
since it appears with the fifth power.


\begin{thebibliography}{99}

\bibitem{Nilles:1982ik}
H.~P.~Nilles,
Phys.\ Lett.\ B {\bf 115} (1982) 193;\\
S.~Ferrara, L.~Girardello, H.~P.~Nilles,
Phys.\ Lett.\ B {\bf 125} (1983) 457;\\
M.~Dine, R.~Rohm, N.~Seiberg, E.~Witten,
Phys.\ Lett.\ B {\bf 156} (1985) 55.

\bibitem{Dine:1985he}
M.~Dine, N.~Seiberg,
Phys.\ Lett.\ B {\bf 162} (1985) 299.

\bibitem{Krasnikov:jj}
N.~V.~Krasnikov,
Phys.\ Lett.\ B {\bf 193} (1987) 37.

\bibitem{Shenker:1990uf}
S.~H.~Shenker,
``The Strength Of Nonperturbative Effects In String Theory,'' RU-90-47,
{\it Workshop on Random Surfaces, Quantum Gravity and Strings, Cargese, 
France, 1990}.

\bibitem{Banks:1994sg}
T.~Banks, M.~Dine,
Phys.\ Rev.\ D {\bf 50} (1994) 7454.

\bibitem{cfx83}
G.~D.~Coughlan, W.~Fischler, E.~W.~Kolb, S.~Raby, G.~G.~Ross,
Phys.\ Lett.\ B {\bf 131} (1983) 59.

\bibitem{deCarlos:1993jw}
T.~Banks, D.~B.~Kaplan, A.~E.~Nelson,
Phys.\ Rev.\ D {\bf 49} (1994) 779;\\
B.~de Carlos, J.~A.~Casas, F.~Quevedo, E.~Roulet,
Phys.\ Lett.\ B {\bf 318} (1993) 447.

\bibitem{Brustein:nk}
R.~Brustein, P.~J.~Steinhardt,
Phys.\ Lett.\ B {\bf 302} (1993) 196.

\bibitem{Dine:2000ds}
M.~Dine,
Phys.\ Lett.\ B {\bf 482} (2000) 213.

\bibitem{Buchmuller:2003is}
W.~Buchm\"uller, K.~Hamaguchi, M.~Ratz,
Phys.\ Lett.\ B {\bf 574} (2003) 156.

\bibitem{Huey:2000jx}
G.~Huey, P.~J.~Steinhardt, B.~A.~Ovrut, D.~Waldram,
Phys.\ Lett.\ B {\bf 476} (2000) 379.

\bibitem{Barreiro:2000pf}
T.~Barreiro, B.~de Carlos, N.~J.~Nunes,
Phys.\ Lett.\ B {\bf 497} (2001) 136.

\bibitem{kap89}
J.~I.~Kapusta,
``Finite Temperature Field Theory,'' Cambridge, 1989.

\bibitem{Kajantie:2002wa}
K.~Kajantie, M.~Laine, K.~Rummukainen, Y.~Schr{\"o}der,
Phys.\ Rev.\ D {\bf 67} (2003) 105008.

\bibitem{Arnold:ps}
P.~Arnold, C.~X.~Zhai,
Phys.\ Rev.\ D {\bf 50} (1994) 7603;
Phys.\ Rev.\ D {\bf 51} (1995) 1906;\\
C.~X.~Zhai, B.~Kastening,
Phys.\ Rev.\ D {\bf 52} (1995) 7232.

\bibitem{deCarlos:1992da}
B.~de Carlos, J.~A.~Casas, C.~Mu\~noz,
Nucl.\ Phys.\ B {\bf 399} (1993) 623.

\bibitem{Casas:1996zi}
J.~A.~Casas,
Phys.\ Lett.\ B {\bf 384} (1996) 103.

\bibitem{Binetruy:1996nx}
P.~Binetruy, M.~K.~Gaillard, Y.~Y.~Wu,
Nucl.\ Phys.\ B {\bf 493} (1997) 27;
Phys.\ Lett.\ B {\bf 412} (1997) 288.

\bibitem{Barreiro:1997rp}
T.~Barreiro, B.~de Carlos, E.~J.~Copeland,
Phys.\ Rev.\ D {\bf 57} (1998) 7354.

\bibitem{gravitino}
M.~Y.~Khlopov, A.~D.~Linde,
Phys.\ Lett.\ B {\bf 138} (1984) 265;\\
J.~R.~Ellis, J.~E.~Kim, D.~V.~Nanopoulos,
Phys.\ Lett.\ B {\bf 145} (1984) 181;\\
T.~Moroi, H.~Murayama, M.~Yamaguchi,
Phys.\ Lett.\ B {\bf 303} (1993) 289.

\bibitem{FY}
M.~Fukugita, T.~Yanagida,
Phys.\ Lett.\ B {\bf 174} (1986) 45.

\bibitem{Buchmuller:2004nz}
W.~Buchm{\"u}ller, P.~Di~Bari, M.~Pl{\"u}macher, 
arXiv:hep-ph/0401240.

\bibitem{Fujii:2002jw}
M.~Fujii, K.~Hamaguchi, T.~Yanagida,
Phys.\ Rev.\ D {\bf 65} (2002) 115012.

\bibitem{Buchmuller:2003gz}
W.~Buchm{\"u}ller, P.~Di Bari, M.~Pl{\"u}macher,
Nucl.\ Phys.\ B {\bf 665} (2003) 445.

\bibitem{Moroi:1999zb}
T.~Moroi, L.~Randall,
Nucl.\ Phys.\ B {\bf 570} (2000) 455.

\bibitem{Kolb-Turner} E.~Kolb, M.~Turner, {\it The Early Universe}
(Addison-Wesley, Redwood City, CA, 1990).

\bibitem{DineRandallThomas}
M.~Dine, L.~Randall, S.~Thomas,
Phys.\ Rev.\ Lett.\  {\bf 75} (1995) 398.

\bibitem{curvaton}
D.~H.~Lyth, D.~Wands,
Phys.\ Lett.\ B {\bf 524} (2002) 5;\\
T.~Moroi, T.~Takahashi,
Phys.\ Lett.\ B {\bf 522} (2001) 215
[Erratum-ibid.\ B {\bf 539} (2002) 303].

\bibitem{Asaka:1999jb}
T.~Asaka, K.~Hamaguchi, M.~Kawasaki, T.~Yanagida,
Phys.\ Rev.\ D {\bf 61} (2000) 083512.

\bibitem{gkp02}
S.~B.~Giddings, S.~Kachru, J.~Polchinski,
Phys.\ Rev.\ D {\bf 66} (2002) 106006.

\bibitem{kkx03}
S.~Kachru, R.~Kallosh, A.~Linde, S.~P.~Trivedi,
Phys.\ Rev.\ D {\bf 68} (2003) 046005.

\bibitem{ggx03}
M.~Gra\~na, T.~W.~Grimm, H.~Jockers, J.~Louis,
arXiv:hep-th/0312232.

\end{thebibliography}
\end{document}